\documentclass[12pt]{article}
\usepackage{authblk}
\usepackage{epsfig,times,latexsym,enumerate,lscape,flafter,amsmath,amssymb,
algorithm,algpseudocode,marginnote}
\begin{document}

%\begin{frontmatter}



\title{A Communication Efficient and Scalable Distributed Data Mining for the Astronomical Data}


\author{Aruna Govada\thanks{garuna@goa.bits-pilani.ac.in} \hspace{0.1pt} and Sanjay K. Sahay\thanks{ssahay@goa.bits-pilani.ac.in}} 
%\ead{ssahay@goa.bits-pilani.ac.in}

%\address{\small Department of Computer Science and Information System, Birla Institute
%of Technology and Science, K. K. Birla Goa Campus, Goa-403726,India}
\affil{\small Department of Computer Science and Information System, BITS, Pilani, K. K. Birla Goa Campus, NH-17B, By Pass Road, Zuarinagar-403726, Goa, India}
\date{}
\maketitle

\vspace{-0.40in}

\begin{abstract}

In 2020, $\sim$ 60PB of archived data will be accessible to the astronomers. But to analyze such a paramount data will be a challenging task. This is basically due to the computational model used to download the data from complex geographically distributed archives to a central site and then analyzing it in the local systems. Because the data has to be downloaded to the central site,  the network BW limitation will be a hindrance for the scientific discoveries. Also analyzing this PB-scale on local machines in a centralized manner is challenging.
In this virtual observatory is a step towards this problem, however, it does not provide the data mining model. Adding the distributed data mining layer to the VO can be the solution in which the knowledge can be downloaded by the astronomers instead the raw data and thereafter astronomers can either reconstruct the data back from the downloaded knowledge or use the knowledge directly for further analysis.Therefore, in this paper, we present Distributed Load Balancing Principal Component Analysis for optimally distributing the computation among the available nodes to minimize the transmission cost and downloading cost for the end user. The experimental analysis is done with Fundamental Plane(FP) data, Gadotti data and complex Mfeat data. In terms of transmission cost, our approach performs better than Qi. et al. and Yue.et al. The analysis shows that with the complex Mfeat data $\sim$ 90\% downloading cost can be reduced for the end user with the negligible loss in accuracy.

\end{abstract}

\indent\indent{\bf Keyword}: {\small\it Distributed Data Mining, Astronomical Data, PCA, Load Balancing}

%\end{keyword}

%\end{frontmatter}

\section{Introduction}
\par Astronomy is afloat with data and an estimate shows that by 2020 more than 60 PB of archived data will be electronically accessible to astronomers. But the complete analysis of the whole accumulated data distributed globally is challenging, not only due to the volume of data but also the communication cost. In general, the computational model used in astronomy is to download the data from archives to a central site and is analyzed in the local machines. Hence, the network bandwidth limitations will be a hindrance for scientific discoveries and even analyzing this PB-scale on local machines in a centralized manner is challenging [1][2][3][28]. Understanding the data collection rate ( e.g. Large Synoptic Survey Telescope (LSST) will generate $\sim$ 30 tera bytes of data every night [4]), the centralize technique will not suffice for comprehensive co-analysis to exploit the potential of  distributed archived data [5]. In this Virtual Observatory (VO) is a step towards the problem, however, it does not provide the data mining model [6]. In this adding the distributed data mining layer to the VO can be the solution [12] in which the knowledge can be downloaded by the astronomers instead the raw data, and thereafter astronomers can either reconstruct the data back from the downloaded knowledge or use the knowledge directly for further analysis. 

\par Astronomical data are mostly high dimensions. Hence, reducing the dimensions of interrelated data will reduce the downloading cost of end users. To reduce the dimension of 
data, a  technique called principal component analysis (PCA) are used in many fields [8]. It reduces the dimensionality of the data 
set of interrelated variables with retaining the variation present in the data [9] [10]. The technique is linear as its components are linear combinations of the original variables, but non-linearity is preserved in the dataset. In this paper, we use Distributed Load Balancing Principal Component Analysis (DLPCA) which is a distributed version of normal PCA to reduce  the transmission (cost incur for distributing the data among the computational nodes) and downloading cost significantly from globally distributed observatories. The algorithm is scalable and also optimally distribute the computational load among the available resources.

\par For the experimental analysis, we took the Fundamental plane (FP), Gadotti and complex Mfeat datasets and use Java Agent Development Framework (JADE), which simplifies the implementation of multi-agent systems through a middle-ware and complies the Foundation for Intelligent Physical Agents specifications. The experimental analysis of our approach is done by creating the multiple agents. The local principal components are computed and communicated among them for computing the global principal component.

\par The paper is organized as follows. In next section, we
discuss the related work on the analysis of globally distributed astronomical data. 
Section 3 briefly describe the mathematics of the PCA to provide an intuitive feeling of it.
In section 4 we present our algorithm for the communication efficient and scalable distributed data mining using DLPCA. In section 5 we present the experimental results. Section 6 discusses the computational cost, load balancing and scalability of our approach. Finally, section 7 contains the conclusion and future direction of the paper.

\section{Related Work}

Big data analysis are primarily based on Distribute Data Mining (DDM) to processes the heterogeneous data from databases located at different places. In literature, various DDM techniques have been proposed for the analysis of heterogeneous data sets. These techniques differs from the centralized data mining in which the analysis is done after downloading the data to one single location. 
Distributed Data Mining based on PCA can be done in two ways; either distributing the data horizontally or vertically [11][12]. A distributed PCA algorithm based on the integration of local covariance matrices for the distributed databases which are horizontally partitioned is given by Qi et.al [13]. If the data is vertically distributed then it is necessary that all the considered sites are associated with a unique way of matching the distributed data [12], for e.g in astronomy, it is done using right ascension and declination (RA, DEC) of the objects. A randomized PCA is discussed by  Nathan Halko et al. [14], for the datasets which are too large to store in the Random Access Memory(RAM).

 \par Astronomical research communities do data mining for large datasets e.g. F-MASS [15], the Auton Astro-statistics Projects [16]. However, this project does not fully based on Distribute Data Mining.  A project called Grid Based Data Mining for Astronomy (GRIST) [17] was the first attempts for large scale data mining in astronomy. Projects in Virtual Observatories such as Japanese Virtual Observatory (JVO), US National Virtual Observatory (NVO), European Virtual Observatory (EURO-VO) and International Virtual Observatory (IVOA), basically integrate and federate archive systems dispersed in a Grid by standardizing XML schema, data access layer, and query language of archival data [17]. In this NVO has developed an information technology infrastructure enabling easy and robust access to distributed astronomical archives, from which users can search and gather data from multiple archives with basic statistical analysis and visualization functions. Giannella et.al [12] describe the architecture of a system called Distributed Exploration of Massive Astronomy Catalogs (DEMAC) for distributed data mining of large astronomical catalogs. The system is designed to sit on top of the existing national virtual observatory environment to provide tools for distributed data mining without downloading the data to a centralized server. Srivastava et. al [18] proposed a distributed and multi-threaded Automated Hierarchical Density  Data in Astronomy: From the Pipeline to the Virtual Observatory clustering algorithm to produce computationally efficient high-quality clusters and scalable from 1-1024 compute-cores. For massive astronomical data analysis distributed CPU/GPU architecture is proposed to handle the data in peta scales [19]. Recently a cloud based data mining system CANFAR+Skytree is proposed at Canadian Astronomy Data Centre [20]. 
Kargupata et. al [21] proposed the solutions for distributed clustering using collective principal component analysis.  Their work is  mainly focused to obtain a good estimation of the global covariance matrix with a trade-off between communication cost and information loss. Further, Yue. et. al [22] proposed a better DDM than Kargupata [21] in which one can achieve a better accuracy with the same communication cost. Our proposed algorithm is communication efficient and scalable DDM based on PCA to further reduce the communication cost with better accuracy which neither needs  to send the local datasets to a central site nor require to reconstruct the local data for calculating the global principal components. 

\section{Principal Component Analysis}

Principal component analysis is a simple non-parametric method to reduce the dimension of the datasets of possibly correlated variables into a smaller number of uncorrelated variables. The first component has maximum variability and each remaining components contains rest of the variabilities. It is abundantly used in many fields viz. astronomy, computer graphics etc. In this section, we briefly describe the mathematics of the PCA to provide an intuitive feeling of it. For details mathematical illustrations the springer series in statistics ``Principal Component Analysis 2e" by I.T. Jollife is a good source [23].
Suppose we have a dataset as
$$[\mathbf{X}]_{n \times m} =  (X_o, X_1, X_2, X_3, ......... X_{l-1})$$
where $X_j$ is a $n \times m_j$ matrix and $m_j$ is the number of columns in $X_j.$

The covariance matrix of the data can be computed as

$$ cov_j^{pq} = \frac{\sum_{i = 1}^{i = n}  (X_{j_i}^p -\mu_j^p) (X_{j_i}^q -\mu_j^q)}{n - 1}$$

where, $\mu_j^p, \mu_j^q$ is the mean of the $p_{th}$ and $q_{th}$ column of the $X_j$ matrix.

The obtained covariance matrix will be a square matrix and symmetric. If the two columns data are completely uncorrelated, then the covariance will be zero. However, there may be a non-linear dependency between two variables that have zero covariance. As covariance matrix is symmetric, hence its eigenvalues and eigenvectors can be obtained by solving the equations
$$ cov_j^{pq}E = \lambda; \quad |cov_j^{pq} -\lambda I | = 0$$

where, $E$ is the eigenvector of eigenvalue $\lambda$ and $I$ is the identity matrix of the same order of  $ cov_j^{pq}$. Now a matrix $P$ can be made consists of eigenvectors of the covariances matrices. The computed eigenvectors are ordered according to its significance. To construct the reduced dataset the least significant eigenvectors are left out and computed as \\

\noindent Reduced dataset ($n \times l$ matrix) = Original dataset (say $n \times m$ matrix) - mean  $\times$ Reduced eigenvector matrix (say $m \times l$ matrix)\\

\noindent Original dataset can be computed as follows\\

Original dataset ($n \times m$ matrix) = Reduced dataset ($n \times l$ matrix) $\times$ (Reduced eigenvector matrix)$^T$ ($l \times m $ matrix)$+$ original mean. \\ \\
where $ n $ is the number of rows, $ m $ is the number of columns and $l$ is the reduced number of columns.
  
\section{Communication Efficient and Scalable DDM using Distributed Load Balancing PCA}

Our approach is basically a communication efficient and scalable DDM for the analysis of astronomical data . The algorithm is described below in eight steps. 

\begin{enumerate}[i)]
\item Let the data is vertically partitioned among the $l$-sites as
$$[\mathbf{X}]_{n \times m} =  (X_o, X_1, X_2, X_3, ......... X_{l-1})$$
where data $X_j$ is a $n \times m_j$ matrix resides at the site $S_j$ and $m = \sum_{j = 0}^{l -1} m_j$.  

\item Normalize locally all the columns data of each sites $S_j$.

\item Compute the covariance between all the columns of every considered sites, represented as

$$ cov_j^{pq} = cov(S_j^p,S_j^q);\;\; p \ne q = 1,2,3,....m_j$$

where $m_j$ is the number of columns of the site $j$.\\

\item Locally find the eigenvectors and eigenvalues from the covariance matrix of each site.

\item Compute the projected data from the dominant local principal components of all sites $S_j$ and send the data as follows

\begin{itemize}
\item If the total number of sites is even, say $2r; r \ge 1$, then send each $S_j$ column data to $S_k$, where

\begin{itemize}
\item  $ k = j + s$ for all $0 \le j \le (r - 1)$ and $1 \le s \le r$, and
\item  $ k = (j + s)\% 2r$, for all $r \le j \le (2r - 1)$ and $1 \le s \le (r-1)$.

\end{itemize}

\item If the total number of sites is odd, say $2r + 1; r \ge 1$, then send $S_j$ to $S_k$, where
\begin{itemize}
\item  $ k = (j + s) \% (2r + 1)$ for all $0 \le j \le 2r$ and $1 \le s \le r$.
\end{itemize}

\end{itemize}

The above steps optimally balance the computational load among the available nodes. For eg. Fig. 1 \& Fig. 2 depict for even and odd number of nodes respectively.
\item Compute global covariance matrix from the projected data as follows

 $$cov_{jk}^{uv} = cov(S_j^u,S_k^v); \;\; u = 1,2,3,...m_j^r; \; v=1,2,3,...m_k^r \; \;\;j \ne k$$.

where $m_j^r$ and $m_k^r$ are the reduced number of columns in the $j_{th}$ and $k_{th}$ site respectively.

\item Using the global eigenvectors project the data on global principal component axis.

\item Now user can download the global eigenvectors, local eigenvectors and the computed global dominant projected data  and can reconstruct the data from it for the scientific discoveries as discussed in the previous section.

\end{enumerate}

\begin{figure}[!htb]
    \centering
    \begin{minipage}[t]{0.49\textwidth}
        \centering
        \includegraphics[width=0.9\linewidth, height=0.25\textheight]{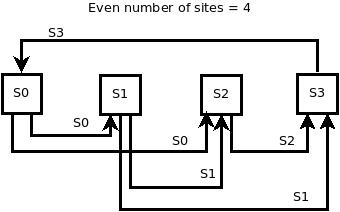}
        \caption{\small \sl Load Balancing for even no.of sites} 
\label{fig:mno}
    \end{minipage}
    \begin{minipage}[t]{0.49\textwidth}
        \centering
        \includegraphics[width=0.9\linewidth, height=0.25\textheight]{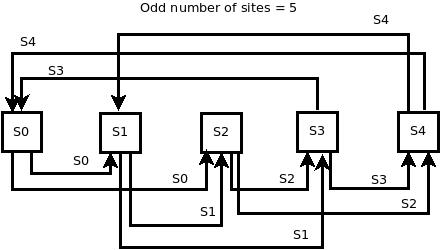} 
\caption{\small \sl Load Balancing for odd no.of sites.}
\label{fig:profilec_mal} 
    \end{minipage}
     \end{figure}

\begin{algorithm}
\caption {DLPCA}
\textbf{INPUT: } Data $X_j$ of all the sites $S_j$ \\
\textbf{OUTPUT: } Global PC's
\begin{algorithmic}[1]

\For{ each site $j$ compute the PC's }

\State Compute $\mu_j$ mean of all columns of $X_j$ data 

\State Compute the covariance matrix $cov_j^{pq} = \frac{\sum_{i = 1}^{i = n}  (X_{j_i}^p -\mu_j^p) (X_{j_i}^q -\mu_j^q)}{n - 1}$ where, $\mu_j^p, \mu_j^q$ is the mean of the $p_{th}$ and $q_{th}$ column of the the $X_j$ matrix.

\State Compute Eigenvectors  $cov_j^{pq}E_l = \lambda; \quad |cov_j^{pq} -\lambda I | = 0$ where, $E_l$ is the eigenvector of eigenvalue $\lambda$ and $I$ is the identity matrix of the same order of  $ cov_j^{pq}$

\State  Compute Principal components $PC_j$ = [ $[E_j]^T$ * $[X_j]^T$ ] $^T$ 
\EndFor
\If { the number of sites is even say $2r, r \ge 1$}
\For{j=0 to (r-1)}
\For{ s= 1 to r }
\State k=j+s and Send $PC_j$ of $S_j$ to $S_k$
\EndFor
\EndFor
\For{j=r to (2r-1)}
\For{s=1 to (r-1)}
\State k=(j+s)\%2r and Send $PC_j$ of $S_j$ to $S_k$ 
\EndFor
\EndFor
\EndIf
\If{the number of sites is odd say $2r+1 ,r \ge 1$}
\For{j=0 to 2r}
\For{s=1 to r}
\State k=(j+s)\%(2r+1) and Send $PC_j$ of $S_j$ to $S_k$
\EndFor
\EndFor
\EndIf
\State Compute the cross covariances   $$cov_{jk}^{uv} = cov(S_j^u,S_k^v); \;\; u = 1,2,3,...m_j^r; \; v=1,2,3,...m_k^r \; \;\;j \ne k$$. and then Global Covariance matrix $covE_G$
\State Compute the global Eigenvectors  $covE_G = \lambda; \quad |cov_G -\lambda I | = 0$ where, $E_G$ is the eigenvector of eigenvalue $\lambda$ and $I$ is the identity matrix of the same order of  $ cov_G$ and project the data on Global PC's
  
\end {algorithmic}

\end{algorithm}

\section{Experimental Results}
Experimental analysis of the proposed algorithm are done with the fundamental plane data (3 columns and 224 rows) [24], gadotti data (7 columns and 946 rows) [25]  and the complex mfeat data (distributed in 6 files having 649 columns and 2000 rows)  [26]. The analysis is focused on two major aspects of the distributed computational environment:
\begin {itemize}
\item  Reduction in transmission cost among the computational nodes by wisely applying the concept of PCA and the results are compared with   Qe et. al and Yue et. al [13] [22]
\item Reduction in downloading cost to the end user, so that the hindrance of the network bandwidth can be minimized for the scientific discoveries.
\end {itemize}

\subsection{Fundamental Plane Data}

 \par In astronomy finding the correlation between the observed quantities plays an important role because it can explain the 
 formations/evolutions of these astronomical objects and can also give a method to measure different quantities. The fundamental plane (FP) is a linear relationship between the effective radius  ($r_e$), the average surface brightness within the effective radius ($\mu_e$) and the velocity dispersion ($\sigma_e$ ) of normal elliptical galaxies. Hence, from the measured quantities viz. $\mu_e$ and $\sigma_e$, one can find the approximated value of  $r_e$, a difficult task in observational astronomy. 
 
 \par To test our approach and the developed code, we recomputed the known FP data [24], by computing all the three PCS and cross verified  with the online IUCAA VO observatory [27] and found that its lie in the same plane (Fig.3). Also, the computed PCs by our method is same as the given in the IUCAA VO observatory. We reconstructed the FP data with two dominant PC and found that it 
 almost same as the original data (Fig.4), hence reduces the downloading cost $\sim$33\%. In subsequent subsection, we discuss the errors between the dominant local PCs and reduced global PCs computed with our algorithm by taking Gadotti data.

 \begin{figure}[!htb]
    \centering
    \begin{minipage}[t]{0.49\textwidth}
        \centering
        \includegraphics[width=0.95\linewidth, height=0.4\textheight]{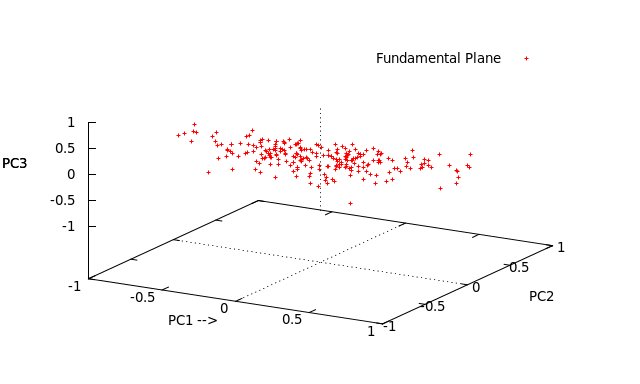}
        \caption{\small \sl All three PCs of FP data lie in same plane.} 
\label{fig:mno}
    \end{minipage}
    \begin{minipage}[t]{0.49\textwidth}
        \centering
        \includegraphics[width=0.95\linewidth, height=0.3\textheight]{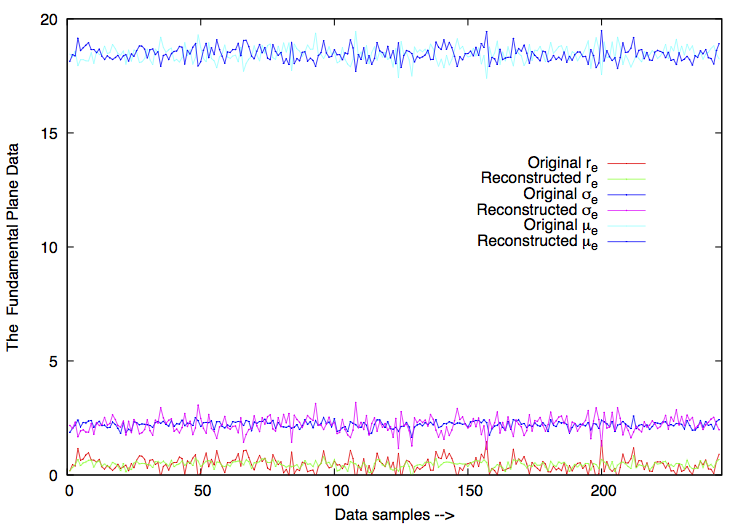} 
\caption{\small \sl Comparison of original data and reconstructed data with two dominant PCs of the FP data.}
\label{fig:profilec_mal} 
    \end{minipage}
     \end{figure}

\subsection{Gadotti Data}

To study the error in the global PCs and the reduction in the downloading cost for the end users, we took Gadotti data which consists of 930 rows and seven columns. For the analysis, we computed the global PCs from the different combination of the considered local PCs.(Table 1)
The estimated error between the actual global PCs and computed global PCs by DLPCA are shown in figure (Fig.5). We found that the error in the global PCs reduces significantly and after (2,3) combination the error is negligible.
[$\sim 10^{-2}\%$ for (2,3) and $\sim 10^{-4}\%$ for $(3,3)$]. Therefore, it will suffice to reconstruct the original data by
taking only the five local dominant PCs which will reduce the downloading cost by $\sim 24\%$. The complete trade-off in error of the taken PCs and the downloading cost is shown in the Fig.5. From analysis we find that the global PC1 error is less compared to other PCs, this is basically due to mean of the first column data is very high compared to other six column data (the  mean of the respective columns data are 20.195, 0.070, 1.484, 0.072, 2.999, 0.473, 0.4728

\begin{table}[ht]
\label{}
\centering
{\begin{tabular}{|c|c|c|}
\hline
No. of Local PCs from $G_{S0}$ & No. of Local PCs from $G_{S1}$ & No. of Global PCs taken for\\
&& the reconstruction of data\\

\hline 
2 (4) & 2 (3)  & 4 \\
\hline
2 (4) & 3 (3) & 5 \\
\hline
3(4)  & 3 (3) & 6 \\
\hline
4 (4) & 3 (3) & 7 \\
\hline
\end{tabular}}
\caption{Global PCs from the different combinations of local PCs.}

\end{table}

\begin{center}
\begin{figure} [!htb]
\includegraphics[scale=0.6]{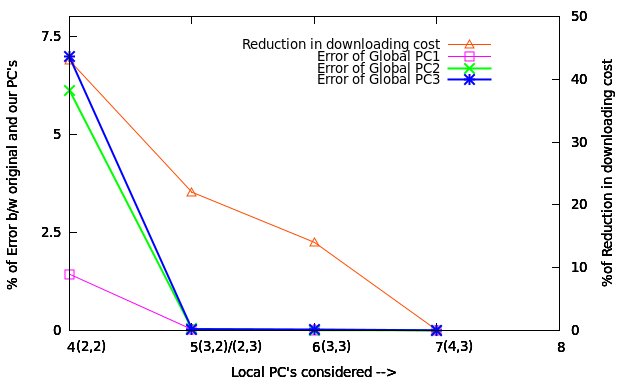} 
\caption{\small \sl Gadotti Difference} 
\end{figure}
\end{center}
     
     \clearpage

\subsection{Mfeat Data}
To study the performance of our approach to a complex data, we took mfeat data which consists of 2000 rows and are distributed in six data files as follows [26]  :
\begin{enumerate}[1.]
\item mfeat-fac: 216 profile correlations;
\item mfeat-fou: 76 Fourier coefficients of the character;
\item mfeat-kar: 64 Karhunen–Love coefficients;
\item mfeat-mor: 6 morphological features;
\item mfeat-pix: 240 pixel averages in 2 x 3 windows;
\item mfeat-zer: 47 Zernike moments.
\end{enumerate}

Following our method described in section 4.1 we computed the transmission cost versus the accuracy i.e. angle between the actual and global dominant PCs (Fig. 12 -15). 
For the purpose, we first computed the PCs variance of all the six datasets (Fig. 6-11) and studied the various combinations of the dominant PCs 
(Table 2) to find the best combination among them based on their variances.

\par From the computed PC variance, we observed that after top 15, 6, 10, 1, 20, 10 PCs of fac, fou, kar, mor, pix, zer respectively the variance in PCs are almost negligible. Hence, following our approach  we calculated the transmission cost versus the angle between the actual and four global dominant PCs and compared with Qi et. al. and Yue. et. al. (Fig. 12-15). The transmission cost among the computational nodes is calculated as follows.

 $$\sum_{j=0}^{j=l-1} (Tr_j)(m_j)(n) $$ \\ where $Tr_j$ is the number of transfers of $\text{j}_{th}$ computational node data to other computational nodes [Fig.1 \& Fig.2] , $m_j$ is the number of reduced columns of the $\text{j}_{th}$ site and $n$ is the number of rows of $\text{j}_{th}$ site which is common to all the sites. Our algorithm outperforms Qi et. al. in the transmission cost where as the accuracy is less than 0.1\% else it is more or less same. However for PC4 transmission cost with the accuracy is always less than Qi.et. al. [13] 

\par If the local PC's are not distributed among the computational nodes then our approach outperforms (Fig. 17-19) Qi et. al. and Yue. et. al. with the exception of PC1 (Fig. 16) when compared with Yue. et. al. The less accuracy compared to Yue. et. al may be due to the complex nature of the data .

\begin{table}[ht]
\label{}
\centering
{\begin{tabular}{|c|c|c|c|c|c|c|}
\hline
fac& fou & kar & mor & pix & zer & Transmission Cost Among the Nodes \\
\hline
1 & 1 & 1 &1 &1 &1 & 3.0$\times 10^4$\\
\hline
4 &1 & 1 & 1 & 4 & 1 & 6.0$\times 10^4$\\
\hline
6 & 2 & 2 & 1 & 6 & 2 & 9.6$\times 10^4$\\
\hline
10 & 4 & 4 & 1 & 10 & 4 & 1.46$\times 10^5$\\
\hline
10 & 4 & 4 & 1 & 13 & 6 & 1.84$\times 10^5$\\
\hline
14 & 5 & 7 & 1 & 14 & 8 & 2.46$\times 10^5$\\
\hline
15 & 6 & 10 & 1 & 20 & 10 & 3.08$\times 10^5$\\
\hline
\end{tabular}}
\caption{mfeat transmission cost}
\end{table}

\begin{figure}[!htb]
    \centering
    \begin{minipage}[t]{0.49\textwidth}
        \centering
        \includegraphics[width=0.95\linewidth, height=0.25\textheight]{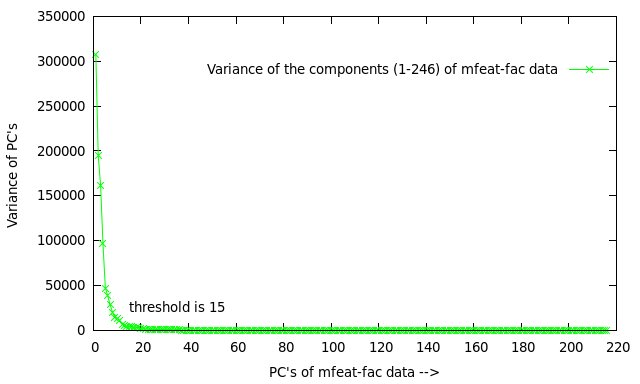}
        \caption{\small \sl Mfeat-fac data PCs variance.} 
\label{fig:mno}
    \end{minipage}
    \begin{minipage}[t]{0.49\textwidth}
        \centering
        \includegraphics[width=0.95\linewidth, height=0.25\textheight]{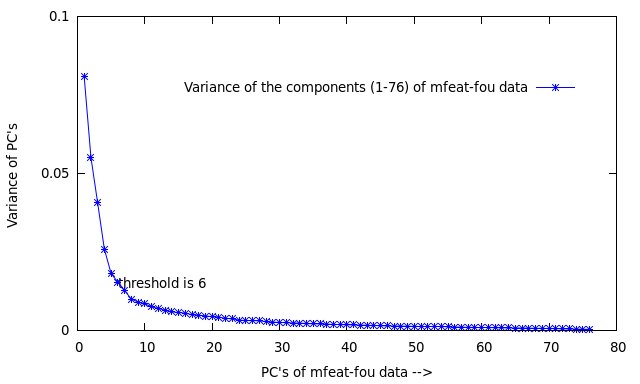} 
\caption{\small \sl Mfeat-fou data PCs variance.}
\label{fig:profilec_mal} 
    \end{minipage}
     \end{figure}

\begin{figure}[!htb]
    \centering
    \begin{minipage}[t]{0.49\textwidth}
        \centering
        \includegraphics[width=0.95\linewidth, height=0.25\textheight]{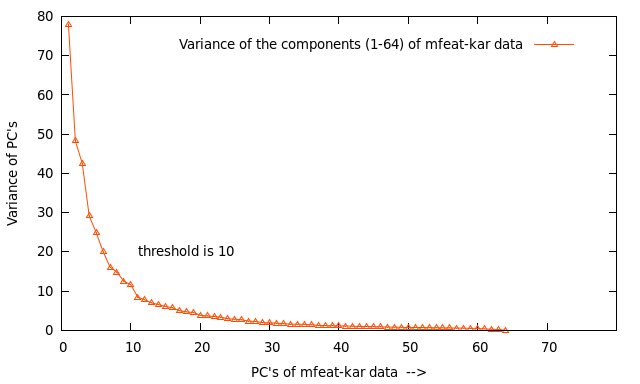}
        \caption{\small \sl Mfeat-kar data PCs variance.} 
\label{fig:mno}
    \end{minipage}
    \begin{minipage}[t]{0.49\textwidth}
        \centering
        \includegraphics[width=0.95\linewidth, height=0.25\textheight]{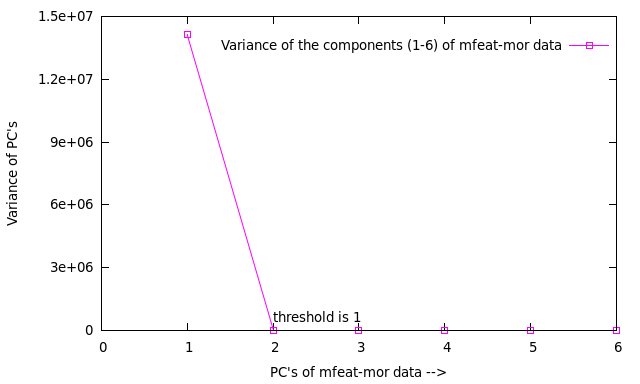} 
\caption{\small \sl Mfeat-mor data PCs variance.}
\label{fig:profilec_mal} 
    \end{minipage}
     \end{figure}

\begin{figure}[!htb]
    \centering
    \begin{minipage}[t]{0.49\textwidth}
        \centering
        \includegraphics[width=0.95\linewidth, height=0.25\textheight]{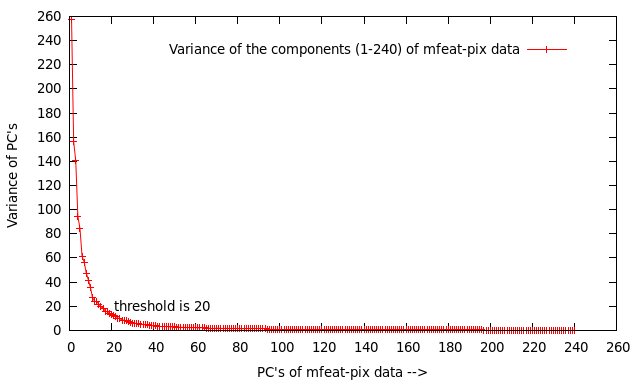}
        \caption{\small \sl Mfeat-pix data PCs variance.} 
\label{fig:mno}
    \end{minipage}
    \begin{minipage}[t]{0.49\textwidth}
        \centering
        \includegraphics[width=0.95\linewidth, height=0.25\textheight]{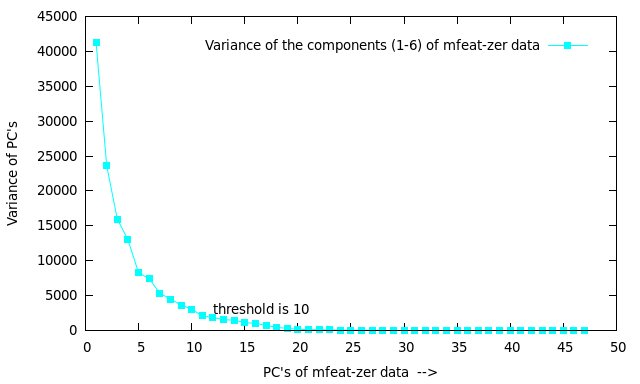} 
\caption{\small \sl Mfeat-zer data PCs variance.}
\label{fig:profilec_mal} 
    \end{minipage}
     \end{figure}

\begin{figure}[!htb]
    \centering
    \begin{minipage}[t]{0.49\textwidth}
        \centering
        \includegraphics[width=0.95\linewidth, height=0.25\textheight]{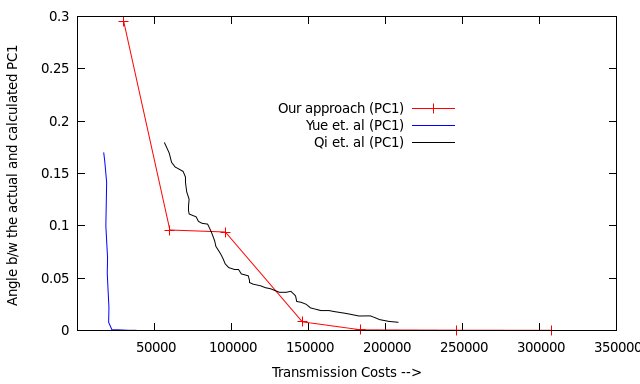}
        \caption{\small \sl Comparison of the transmission cost w.r.t angle between actual and calculated global PC1 with our approach.} 
\label{fig:mno}
    \end{minipage}
    \begin{minipage}[t]{0.49\textwidth}
        \centering
        \includegraphics[width=0.95\linewidth, height=0.25\textheight]{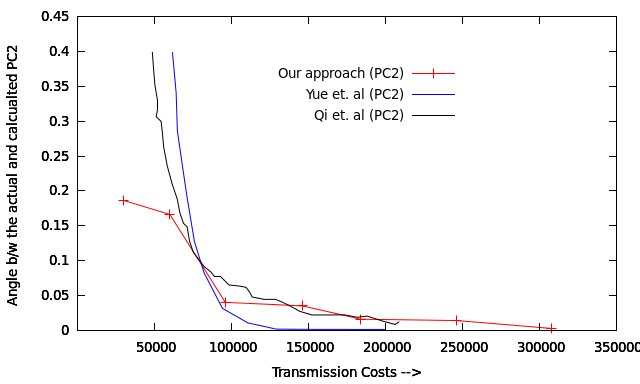} 
\caption{\small \sl Comparison of the transmission cost w.r.t angle between actual and calculated global PC2 with our approach.}
\label{fig:profilec_mal} 
    \end{minipage}
     \end{figure}

\begin{figure}[!htb]
    \centering
    \begin{minipage}[t]{0.49\textwidth}
        \centering
        \includegraphics[width=0.95\linewidth, height=0.25\textheight]{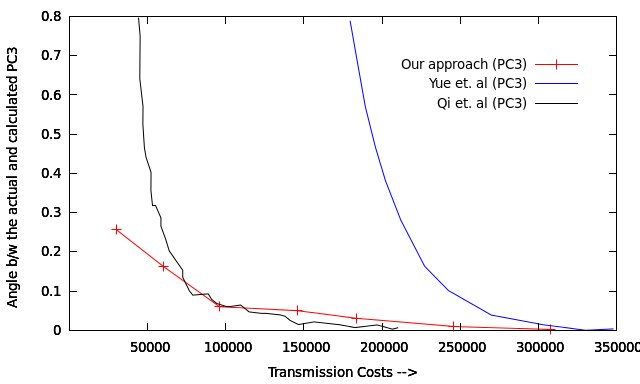}
        \caption{\small \sl Comparison of the transmission cost w.r.t angle between actual and calculated global PC3 with our approach.} 
\label{fig:mno}
    \end{minipage}
    \begin{minipage}[t]{0.49\textwidth}
        \centering
        \includegraphics[width=0.95\linewidth, height=0.25\textheight]{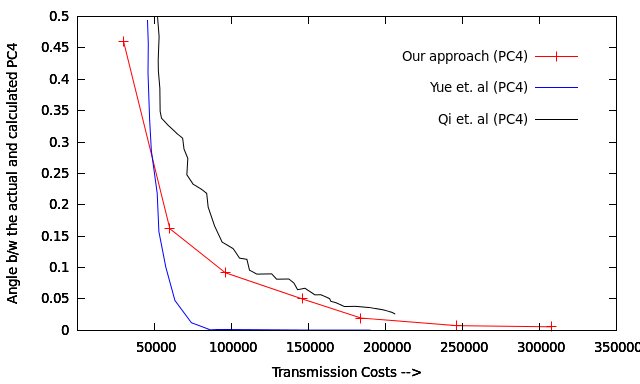} 
\caption{\small \sl Comparison of the transmission cost w.r.t angle between actual and calculated global PC4 with our approach.}
\label{fig:profilec_mal} 
    \end{minipage}
     \end{figure}

\begin{figure}[!htb]
    \centering
    \begin{minipage}[t]{0.49\textwidth}
        \centering
        \includegraphics[width=0.95\linewidth, height=0.25\textheight]{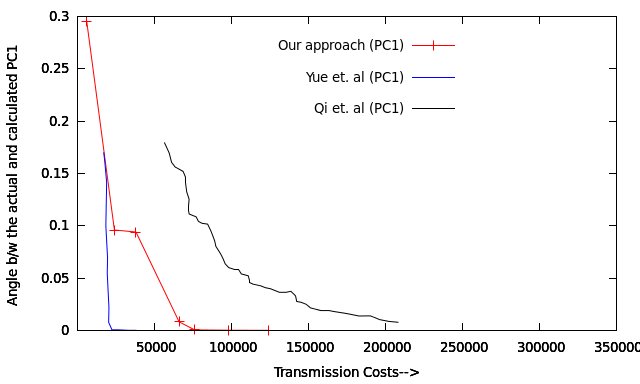}
        \caption{\small \sl Comparison of the transmission cost w.r.t angle between actual and calculated global PC1 by not distributing the load.} 
\label{fig:mno}
    \end{minipage}
    \begin{minipage}[t]{0.49\textwidth}
        \centering
        \includegraphics[width=0.95\linewidth, height=0.25\textheight]{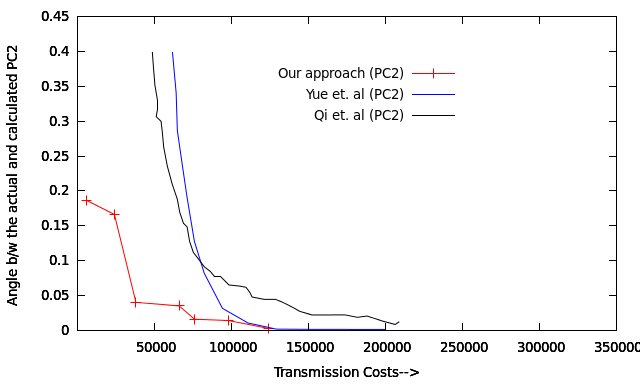} 
\caption{\small \sl Comparison of the transmission cost w.r.t angle between actual and calculated global PC2 by not distributing the load.}
\label{fig:profilec_mal} 
    \end{minipage}
     \end{figure}

\begin{figure}[!htb]
    \centering
    \begin{minipage}[t]{0.49\textwidth}
        \centering
        \includegraphics[width=0.95\linewidth, height=0.25\textheight]{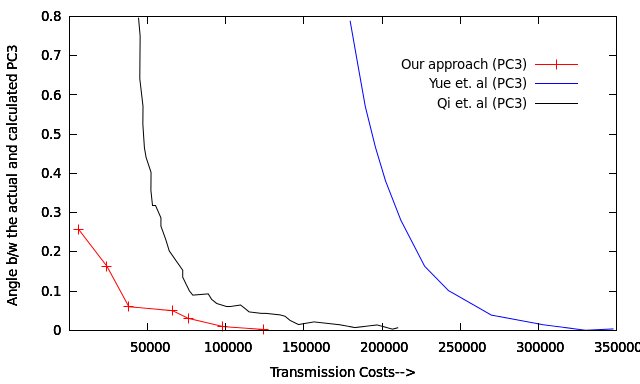}
        \caption{\small \sl Comparison of the transmission cost w.r.t angle between actual and calculated global PC3 by not distributing the load.} 
\label{fig:mno}
    \end{minipage}
    \begin{minipage}[t]{0.49\textwidth}
        \centering
        \includegraphics[width=0.95\linewidth, height=0.25\textheight]{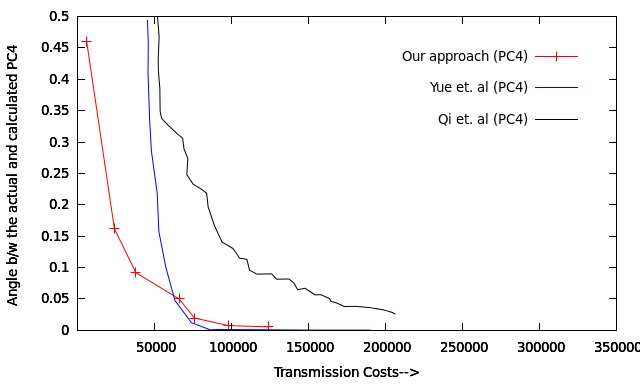} 
\caption{\small \sl Comparison of the transmission cost w.r.t angle between actual and calculated global PC4 by not distributing the load.}
\label{fig:profilec_mal} 
    \end{minipage}
     \end{figure}

\label{}

\section{Computation and Communication}

\subsection{Computational Cost}
We assumed that the data are distributed (vertically partitioned), represented 
as $S_j^{nm_j}$, where $n$ and $m_j$ are the numbers of rows and columns of the $j_{th}$  site. At every site, the first column contains the same source locations (RA, DEC). \\

\par For simplicity, let us assume that all the sites have same computational resources and $T_{cov}$ be the time required to compute the covariance between two columns of any considered site. Therefore, total time requires for computing the covariances between columns  of all the considered sites can be written as

$$\sum_{j=0}^{j=l-1}\frac{m_j(m_j -1)}{2}.T_{cov}$$

\par Now, the total computational 
cost for cross site covariances of all the sites can be given as

$$ m_j^r \times m_k^r .T_{cov} $$

where, $ m_j^r, \;\; m_k^r$ are the reduced number of columns of the $j_{th}$ and $k_{th}$ site.

\bigskip

\par Now the communication/transmission cost among the nodes is
$$\sum_{j=0}^{j=l-1} (Tr_j)(m_j)(n) (T_{com})  $$ \\ where $Tr_j$ is the number of transfers of jth computational node data to other computational nodes [Fig.1 \& Fig.2] , $m_j$ is the number of reduced columns of jth site and $n$ is the number of rows of jth site which is common to all the sites, $T_{com}$ is the communication cost to send one column data from on site to another. 

Therefore, the total computational cost of DLPCA is 

 $$\sum_{j=0}^{j=l-1}\frac{m_j(m_j -1)}{2}.T_{cov} + (m_j^r \times m_k^r) T_{cov} + \sum_{j=0}^{j=l-1} (Tr_j)(m_j)(n) (T_{com}) $$  
 
 by considering $T_{cov} , T_{com} $ as unit cost 

$$\sum_{j=0}^{j=l-1}\frac{m_j(m_j -1)}{2} + (m_j^r \times m_k^r) + \sum_{j=0}^{j=l-1} (Tr_j)(m_j)(n) $$ 

 $$ = \sum_{j=0}^{j=l-1}\frac{m_j(m_j -1)}{2} + (m_j^r \times m_k^r) + K(m_j)(n) $$ 
  where,
  \begin{equation}
   K=
    \begin{cases}
      l(\frac{l}{2}), & \text{if}\; l \; \text{is odd} \\
      \frac{l}{2} (l-1) , & \text{if}\; l \; \text{is even}
    \end{cases}
  \end{equation}

\newpage
\section{Conclusions and Future Direction}

We proposed a communication efficient and scalable DDM using DLPCA for downloading the astronomical data by the end user stored at
different observatories. The algorithm uses distributed load balancing PCA to reduce the transmission cost among the computational nodes and  downloading cost with negligible loss in the information. In our approach the number of transfers of the PCs is half of the number of sites (section 4, fig. 1 and 2) i.e. it is not required that all the sites should have the PCs of all the  other sites but in Qi. et al. the local PCs are sent to one centralized site and then the global PC's are estimated and Yue. et.al. approach does not address the load balance among the sites. 

\par Also, our approach is scalable i.e. one can easily add any number of  new observatory data.The computational load for the computation of cross-site covariances is optimally distributed 
among the computational resources of the observatories. If $m_f$ is the number of final columns 
to be downloaded by the end user then our downloading cost is $O(m_f)$ by neglecting the cost of global and local eigenvectors. The results also show that transmission cost between the computational nodes is less than the approach of Yue et. al [22]. Our experimental test data analysis shows that 
downloading cost will be reduced by $\sim$ 33\% for FP data, $\sim$ 27\% for Gadotti data and $\sim$ 90\% for Mfeat data with a good accuracy of the reconstructed data. The reduction in cost will be more depends on how much end users can afford the loss in information and the volume of data he/she supposes to download. DLPCA is not only applicable to astronomical data but non-astronomical data also. The experimental analysis is done using Fundamental Plane, Gadotti which are astronomical and Mfeat which is non-astronomical. In future, we will do analysis taking in account of latency, bandwidth, processing speed and memory efficiency.  In this, we understand that the local computation of the cross-site covariance can be made more efficient by using alternate computational frameworks like General Purpose Graphics Processing Unit.

\section*{Acknowledgments}
We are thankful to Prof. A.K. Kembhavi, Prof. Dipankar Bhatacharya, IUCAA, Pune and R.K. Roul, Jessica Periera  BITS, Pilani, K.K. Birla Goa Campus for the useful discussions and valuable suggestions. The authors are extremely thankful to the anonymous referee for his hard work in making the detailed suggestions.


\section*{References}

\begin{thebibliography}{00}

\bibitem{} M Weske, M SHacid, C Godart(Eds). Data in Astronomy : From the Pipeline to the Virtual Observatory. {\it WISE 2007 workshops}, LNCS 4832, pp 52-62, 2007. 

\bibitem{} Petr Skoda. Open European Journal on Variable Stars. Nov 2007 pp 32-36, ISSN 1801-5964. 

\bibitem{}  http://www.openskyquery.net/Sky/SkySite/OSQForm/default.aspx.

\bibitem{}  Z Iveric, J A Tyson, E Acosta et al. LSST:from science drivers to reference design and anticipated data products. Version 2.0.9 of June,2011,arxiv:0805.2366.

\bibitem{} Kanishka Bhaduri, Kamalika Das, Kirk Borne et al. 
 Scalable, Asynchronous, Distributed Eigen-Monitoring of Astronomy Data Streams
 {\it Proceedings of the 2009 SIAM International Conference on Data Mining}. pp 247-258.

\bibitem{} Chilingarian, Igor, Bonnarel et al.
   Astronomical Data Analysis Software and Systems XXI, {\it Proceedings of ASP Conference Series 2011}, Vol. 461, {\it Astronomical Society of the Pacific}, 2012, p.307.

\bibitem{}  Yan-Xia Zhang, A-Li Luo, Yong-Heng Zhao.  Outlier detection in astronomical data . {\it In Proc. SPIE 5493, Optimizing Scientific Return for Astronomy through Information Technologies}, 521 (September 16, 2004).

\bibitem{}  I T Joliffe. {\it Principal Component Analysis}. Springer-Verlag, 1986.

\bibitem{}  K. Pearson. {\it On lines and planes of closest fit to systems of points in space}, Phil. Mag., 2 (1901), pp. 559 - 572.

\bibitem{} H Hotelling. {\it Analysis of a complex of statistical variables into principal components}, J. Educ. Psych., 24 (1933), pp. 417 - 441, 498 - 520.

\bibitem{}  H Dutta, C Giannella, K Borne et al. Distributed Top-K Outlier Detection from Astronomy Catalogs using the DEMAC System. {\it In Proceedings of SDM’07}, 2007, pp 473-478.

\bibitem{}  H Dutta and H Kargupta. Distributed Data Mining for Astronomy Catalogs, 2006. {\it The 9th Workshop
on Mining Scientific and Engineering Data Sets} (held in conjunction with SDM 2006).

\bibitem{}  Hairong Qi, Tsei-Wei Wang, J Douglas Birdwell, Global Principal Component Analysis for Dimensionality Reduction in Distributed Data Mining University of Tennessee Knoxville, CRC Press, 2003, p. 324-337.

\bibitem{} Nathan Halko, Per-GUnnar Martisson, Yoel Shkolnisky et al. An algorithm for the principal component analysis of large data sets , SIAM Journal on Scientific Computing, Vol.33 No.5, pp. 2580-2594,2011. 

\bibitem{} The ClassX Project: Classifying the High-Energy Universe. http://heasarc.gsfc.nasa.gov/classx/.

\bibitem{}  The AUTON Project. http://www.autonlab.org/autonweb/19702.html 

\bibitem{} Joseph C Jacob, Daniel S Katz, Craig D Miller et al. Grist: Grid-based Data Mining for Astronomy, Astronomical Data Analysis Software and Systems XIV, {\it ASP Conference Series}, Vol. XXX, 2005.

\bibitem{} Srivatsava Daruru, Sankari Dhandapani, Gunjan Gupta et al. Distributed, Scalable Clustering for Detecting Halos in Terascale Astronomy Datasets. {\it ICDM Workshops 2010}: 138-147.

\bibitem{} A H Hassan, Christopher J Fluke, David G Barnes. Unleashing the Power of Distributed CPU/GPU Architectures: Massive Astronomical Data Analysis and Visualization case study. CoRR abs/1111.6661, 2011.

\bibitem{} Ball N M. CANFAR+Skytree: A Cloud Computing and Data Mining System for Astronomy, Astronomical Data Analysis Software and Systems XXII, 2013, Volume Number :475 p.391 - 394.

\bibitem{}  H Kargupta, W Huang, K Sivakumar et al. Distributed Clustering Using Collective Principal Component Analysis. Knowledge and Information Systems, 3(4): 422 - 448, 2001.

\bibitem{} Yue-Fei Guo,Xiaodong Lin,Zhouu Teng, Xiangyang Xue,Jianping Fan,A coariance-free iterative algorithm for distributed principal component analysis on vertically partitioned data, Pattern Recognition (3) : 1211 - 1219 (12).

\bibitem{} Jolliffe I.T. {\it Principal Component Analysis}, Springer Series in Statistics, Springer, 2nd edition, 2002.

\bibitem{} Jorgensen, Inger; Franx, Marijn; Kjaergaard, Per {\it The Fundamental Plane for cluster E and S0 galaxies }, Monthly Notices of the Royal Astronomical Society, Volume 280, Issue 1, pp. 167-185

\bibitem{} Gadotti, Dimitri A. {\it  Structural properties of pseudo-bulges, classical bulges and elliptical galaxies: a Sloan Digital Sky Survey perspective }, Monthly Notices of the Royal Astronomical Society, Volume 393, Issue 4, pp. 1531-1552.

\bibitem{} Mfeat Data set : https://archive.ics.uci.edu/ml/datasets/Multiple+Features

\bibitem{} IUCAA VO Observatory (Astrostat): http://voi-apps.iucaa.in/astrostat/pages/astrostat.jsp 

\bibitem{} G.Bruce Berriman, Steven L.Groom {\it How will Astronomy Archives Survive the Data Tsunami?} Communications of the ACM Vol.54 Issue 12, Dec 2011 Pages 52-56

\bibitem{} Daniel Grosu, Anthony T. Chronopoulos Ming-Ying Leung 
{\it Load Balancing in Distributed Systems: An Approach Using Cooperative Games} Parallel and Distributed Processing Symposium., Proceedings International, IPDPS 2002 

\end{thebibliography}
\end{document}